\documentclass[12pt,thmsa]{article}
\usepackage{setspace}

\usepackage{graphicx} 
\usepackage{amsthm,amsmath}
\RequirePackage{natbib}
\usepackage{ulem}
\usepackage{bm}
\usepackage{amssymb}
\usepackage{pstricks} 

\newcommand{\I}{\mbox{\bf{I}}}  
\newcommand{\Ind}{\mbox{\bf{1}}}

\bibpunct{(}{)}{;}{a}{}{,}

\begin{document}
\doublespacing
\author{Richard J. Barker\thanks{
Department of Mathematics and Statistics, University of Otago, P.O. Box 56 Dunedin, New Zealand}\; and William A. Link%
\thanks{%
Patuxent Wildlife Research Center, Laurel, MD 20708, USA}}
\title{\textbf{Posterior Model Probabilities Computed From Model-Specific Gibbs Output}}
\maketitle

\hrule
\vspace{0.5cm}
Reversible jump Markov chain Monte Carlo (RJMCMC) extends ordinary MCMC methods for use in Bayesian multimodel inference.  We show that RJMCMC can be implemented as Gibbs sampling with alternating updates of a model indicator and a vector-valued ``palette'' of parameters denoted $\bm \psi$.  Like an artist uses the palette to mix dabs of color for specific needs, we create model-specific parameters from the set available in $\bm \psi$.  This description not only removes some of the mystery of RJMCMC, but also provides a basis for fitting models one at a time using ordinary MCMC and computing model weights or Bayes factors by post-processing the Monte Carlo output.
We illustrate our procedure using several examples.
\\ \\

\vspace{0.5cm}
\hrule

\section{Introduction}
A natural Bayesian approach to problems of multimodel inference is to compute posterior model probabilities, or equivalently Bayes factors, given priors and data.  Bayes factors involve marginal likelihoods and can be difficult to calculate.  Here we address the problem of estimating posterior model probabilities, and provide a representation of reversible jump Markov chain Monte Carlo (RJMCMC) that allows us to use MCMC output obtained 
 fitting models one at a time.  

Techniques have been proposed for computing Bayes factors using MCMC output from independent chains generated for different models \cite[for example,][]{Chib1995} or by using a search over the joint space of model indicators $M \in {\mathcal M}$ and model parameters $\theta_j \in \Theta_j$ given by ${\mathcal M} \times \prod_{j \in {\mathcal M}} \Theta_j$. Either approach is difficult in practice and it is common for model selection to be based instead on a deviance information criterion (DIC) \cite[]{Spiegelhalter2002}.  
However, there is no theoretical justification for using DIC to produce model weights or Bayes factors.

Reversible jump Markov chain Monte Carlo is an extension of ordinary MCMC, for multimodel inference.  In this broader context, the posterior distribution under investigation describes parameters for a collection of models, rather than a single model; furthermore the posterior distribution describes model uncertainty through weights on a categorical variable we call \textit{Model}.  

A key step in implementing RJMCMC is the specification of bijections describing relationships between the parameters of various models \cite[e.g.,][]{Green1995,Gelman2004}.  RJMCMC is usually described in terms of $\binom{K}{2}$ such bijections where $K$ is the number of models in the model set ${\mathcal M}$.  \cite{Link2010} outlined an alternative formulation of RJMCMC as simple Gibbs sampling, alternating between updating a palette of parameters $\bm \psi$, which is of the same dimension for all models, and 
the categorical variable \textit{Model}.  There are \textit{K} bijections, one relating each model's parameters to the palette $\bm \psi$; the $K \choose 2$ bijections typically described are obtained from these.  Careful construction of the palette and \textit{K} bijections  
allows RJMCMC to be carried out using samples from model-specific posteriors, obtained one model at a time. Here we illustrate this approach and extend \cite{Link2010} by showing  
that moves between models can be written so that they involve a direct draw from a known categorical distribution with all models in the sample space.  This formulation obviates the need for use of a Metropolis-Hastings step that only allows pair-wise comparison of models and can be easier to implement than RJMCMC in its usual incarnations.

\section{A description of RJMCMC}
Suppose that we wish to evaluate the relative support provided by data $\bm y$ to models $M_k$, $k=1, 2,\ldots,K$, these models being fully known except for  parameter vectors $\bm \theta^{(k)}$ for which we have specified priors.

RJMCMC can be expressed as simple Gibbs sampling, with draws alternating between the categorical variable \textit{Model} and a universal vector-valued parameter $\bm \psi$.  We compare $\bm \psi$ to an artist's palette: as the artist combines colors on her palette to produce colors needed for specific applications, so components of $\bm \psi$ are combined to produce model-specific parameters $\bm \theta^{(k)}$.  The important feature of RJMCMC is that the entire palette $\bm \psi$ is updated at each step of the Gibbs sampler (rather than simply those components relating to the present model).

Following \cite{Link2010} the palette of parameters $\bm{\psi}$ is a vector of dimension $d$ greater than or equal to the dimension of the most complex model in the model set. Parameter vector $\bm \theta^{(k)}$ can be recovered from the palette $\bm \psi$ by means of a known (invertible) mapping $g_k(\bm \psi) = \bm{\Theta}^{(k)} = (\bm{\theta}^{(k)}, \bm{u}^{(k)})'$.  Vector $\bm{u}^{(k)}$ is irrelevant to model $M_k$, serving only to match the dimension of $\bm{\Theta}^{(k)}$ and $\bm \psi$, so that $g_k(\cdot)$ can be defined as a bijection.
  Thus if model $M_2$ has parameter space of dimension 7, and $d = 10$, vector $\bm u^{(2)}$ will have dimension 3.
  
Note that the $K \choose 2$ bijections typically required for RJMCMC are induced by our $K$ bijections between the palette and model-specific parameter spaces: for models $M_j$ and $M_k$ we have $g_{jk}(\bm \Theta^{(j)}) \equiv g_k \circ g_j^{-1} (\bm \Theta^{(j)}) = g_k(\bm \psi) = \bm \Theta^{(k)}$.


%
We must specify a prior for $\bm \psi$ and in doing so 
accommodate model specific priors $[\bm \theta^{(k)}| Model = M_k]$, which for simplicity we write as 
$[\bm \theta^{(k)}|M_k]$.  We have 
\[
  [\bm \Theta^{(k)}|M_k] = 
  [\bm \theta^{(k)},\bm u^{(k)}|M_k] 
\]
\[
 = [\bm \theta^{(k)}|M_k]\;[\bm u^{(k)}|\bm \theta^{(k)}, M_k].
\]
All that is needed is a specification of $[\bm u^{(k)}|\bm \theta^{(k)}, M_k]$; given that $\bm u^{(k)}$ has no role in inference, it will be convenient to assume it is conditionally independent of $\bm \theta^{(k)}$, so that $[\bm u^{(k)}| \bm \theta^{(k)}, M_k] = [\bm u^{(k)}| M_k]$.  The specific choice does not matter, except for tuning the RJMCMC algorithm.  From $[\bm \Theta^{(k)}|M_k]$ we obtain $[\bm \psi|M_k]$ using the change of variables theorem  in terms of a prior $f_k(\bm \Theta^{(k)}) = [\bm \theta^{(k)}, \bm u^{(k)}|M_k]$.  The prior on $\bm \psi$ is then 
\[
 [\bm \psi] =\sum_k[\bm \psi | M_k][M_k]
\]
where
\begin{equation}
 [\bm{\psi} | M_k] = f_k\left(g_k(\bm{\psi}) \right)
 \left| \frac{\partial g_k(\bm{\psi})}{\partial  \bm{\psi}}\right|. \label{eq:psigivenM}
\end{equation}

%

Under this formulation, Gibbs sampling consists of cyclical sampling of full conditional distributions, alternating between draws from $[\bm \psi | M_k, \bm y]$ to update $\bm \psi$, and from $[\{ M_1, \ldots, M_K \} | \bm \psi, \bm y]$ to update \textit{Model}.  

\subsection*{Updating $\bm \psi$}
To draw from the full conditional $[\bm \psi | M_k, \bm y]$ it suffices to draw from $[\bm \theta^{(k)} | M_k, \bm y]$ and $[\bm u^{(k)} | M_k]$ and then to apply the inverse transformation $g^{-1}_k(\bm \Theta^{(k)})$ to obtain a draw for $\bm \psi$.  
 
The draw from $[\bm \theta^{(k)} | M_k, \bm y]$ can either be made directly, if the distribution is of convenient form, or by simulation.  Another possibility, often an attractive alternative, is to take a random draw from the stored MCMC output of an earlier analysis of model $M_k$.

\subsection*{Updating the model}
The full-conditional for \textit{Model} is categorical with probabilities:
    \[
       \Pr(M_k\;|\;\cdot) = \frac{[\bm y\;|\;\bm \psi, M_k][\bm \psi\;|\;M_k][M_k]}{\sum_j [\bm y\;|\;\bm \psi, M_j][\bm \psi\;|\;M_j][M_j]}\;,
    \]
    for $k=1, \ldots, K$.  If we are willing to calculate all of these probabilities, we can update \textit{Model} by a direct draw from this full-conditional distribution.  Chain means of model indicators $I(Model = M_k)$ converge to the full conditional model probabilities, but greater efficiency is available by using chain means of the full conditional model probabilities, which also converge to the posterior model probabilities.
    
    As an alternative, we can update model indicators by a Metropolis-Hastings step if the model candidate generator only allows limited transitions, for example, to a near neighbor in a graphical model sense. The advantage of this approach is that we compute a smaller set of categorical probabilities, corresponding to the neighborhood set. In this case we must compute posterior model probabilities by chain means of indicators $I(Model = M_k)$.\\
    
Expressing RJMCMC as simple Gibbs sampling provides the key innovation of our formulation: it allows us to fit models one at a time using ordinary MCMC and then compute model weights or Bayes factors by post-processing the Monte Carlo output. Thus, we have a simple 2-stage procedure that can be used for computing model probabilities:

\subsubsection*{Stage 1: Produce samples of $[\bm \psi|\bm y, M_k]$ for each $k$.}  
Begin by sampling $[\bm \theta^{(k)}|\bm y, M_k]$.  This can be accomplished by running an MCMC sampler for $M_k$, processing it in the usual way, discarding any initial burn-in iterations, and storing the results. (In cases where the posterior distribution for $\bm \theta^{(k)}$ is of a known and easily sampled form, we do so.)  For each sampled value $\bm \theta^{(k)}$, independently sample an auxiliary variable $\bm u^{(k)}$ from $[\bm u^{(k)}|M_k]$ and calculate $\bm \psi = g_k^{-1}((\bm \theta^{(k)}, \bm u^{(k)})')$.  The collection of sampled values $\bm \psi$ is a sample of $[\bm \psi|\bm y, M_k]$.
\\
\subsubsection*{Stage 2: Post-process the model specific outputs.}
Posterior model probabilities can be computed in one of two ways.  The first method is based on generating a Markov chain of 
the categorical variable \textit{Model}
 that can be used as a posterior sample from 
$[Model |\bm y]$. 
The second method is based on generating a Markov chain of between-model transition probabilities that can be used to estimate the between-model transition matrix.  The steady state marginal distribution of this matrix corresponds to the posterior model probabilities.\\

{\bf Method 1}
A posterior 
sample \textit{Model}$^{(1)}$, \textit{Model}$^{(2)}$, \ldots, \textit{Model}$^{(j)}$, \ldots\
can be generated as follows:
\begin{itemize}
  \item[(a)] Initialize
  \textit{Model}, say with $Model^{(1)} = M_1$.
  \item[(b)] Iterate from $j=1$ to some large number $J$, and at each step:
  \begin{itemize}
    \item[(i)] If \textit{Model}$^{(j)} = M_k$, draw a value $\bm \psi^{(j)}$ from the stored sample of $[\bm \psi|\bm y, M_k]$ from Stage 1. 
    \item[(ii)] Compute $\pi_k^{(j)} = \Pr(Model = M_k|\bm \psi^{(j)}, \bm y)$, for each $k$.  This calculation requires the Jacobian of the transformation $\bm \Theta^{(k)} = g_k(\bm \psi)$ as in eq. (\ref{eq:psigivenM}), evaluated at $\bm \psi^{(j)}$. 
    \item[(iii)] Sample \textit{Model}$^{(j+1)}$ from a categorical distribution with sample space $\{M_1, M_2,\ldots, M_K\}$ and probability vector $\bm \pi^{(j)}$ =\\ $\left (\pi_1^{(j)},\pi_2^{(j)},\ldots,\pi_K^{(j)} \right )'$. 
  \end{itemize}
\end{itemize}
The relative frequency with which $Model^{(j)} = M_k$ approximates the posterior model probability for $M_k$.  A better estimate (Rao-Blackwellized) is the chain mean of values $\pi_k^{(j)}$. \\

{\bf Method 2}  A further improvement on this approximation can be made by marginalizing at stage 1, obviating the need for the construction of a Markov chain of model indicators in the second stage.  For each model we can compute $\Pr(M_k | \bm \psi, M_h)$ forming a Markov chain of transition probabilities from model $h$ to model $k$ $(h,k \in 1,\ldots, K)$.  These can then be averaged to form an approximation to the stochastic matrix governing model-to-model transitions.  Given an estimate of this transition matrix we can obtain corresponding estimates of the posterior model probabilities as the limiting distribution obtainable by normalizing the left eigenvector of the transition matrix associated with the eigenvalue 1.0 \cite[]{Seber2008}. 

\section{Examples}
\subsection{Radiata pine data}
\cite{Carlin1995}, \cite{Han2001}, and many others analyze data taken from \cite{Williams1959}.  The response variable $y_i$ is the maximum compressive strength parallel to the grain for 42 radiata pine boards.
Two explanatory variables are considered: the first is the specimen's density, $x_i$, and the second is the specimen's density having adjusted for resin content, $z_i$.  Resin increases the density of boards without increasing their compressive strength.  \cite{Carlin1995} considered two models:
\begin{eqnarray*}
 \mbox{Model 1:} & y_i = \alpha+\beta (x_i-\bar{x}) + \varepsilon_i, & \varepsilon_i \sim N(0, \sigma^2_x)
\end{eqnarray*}
and
\begin{eqnarray*}
 \mbox{Model 2:} & y_i = \gamma+\delta (z_i-\bar{z}) + \varepsilon_i, & \varepsilon_i \sim N(0, \sigma^2_z).
\end{eqnarray*}
In both cases the errors $\varepsilon$ are assumed iid among observations, conditional on the parameters. 

As priors, \cite{Carlin1995} used $N((3000, 185)', \mbox{diag}(10^6, 10^4))$ priors on $(\alpha, \beta)'$ and $(\gamma, \delta)'$, and inverse gamma priors on $\sigma^2_x$ and $\sigma^2_z$, both having mean and standard deviation equal to $300^2$.  This quirky choice of priors was made to be vague but with expectations corresponding roughly to the parameter estimates obtained by fitting the model by least squares.

We fitted each of these models independently using BUGS \cite[]{Lunn2000} and the above priors, running three chains of 60,000 each with distinct starting values.  Discarding the first 10,000 of each chain as a burn-in left us with a posterior sample of 150,000 for $\bm \theta^{(1)}$ and $\bm \theta^{(2)}$.  We coded a reversible jump algorithm in which $g_1(\bm \psi) = (\alpha, \beta, \sigma^2_x)'$ and $g_2(\bm \psi) = (\gamma, \delta, \sigma^2_z)'$.  In this case, $[\bm \psi | M_1] = [\bm \psi | M_2]$, and the model update is based on the relative values of the likelihoods weighted by the model priors $\Pr(M_k)$:
\[
  \Pr(M_k | \bm y, \bm \psi) = \frac{ \Pr(M_k) e^{\frac{-1}{2 \psi_3}\sum_{i=1}^{42}(y_i-\mu_{ik})^2}}{\Pr(M_1) e^{\frac{-1}{2 \psi_3}\sum_{i=1}^{42}(y_i-\mu_{i1})^2} + \Pr(M_2) e^{\frac{-1}{2 \psi_3}\sum_{i=1}^{42}(y_i-\mu_{i2})^2}},
\]
where
\[
  \mu_{ik} = \left \{ \begin{array}{ll}\psi_1+\psi_2(x_i-\bar{x}) & k = 1\\ \psi_1+\psi_2(z_i-\bar{z}) & k = 2 \end{array} \right. .
\]

\begin{figure}[!htbp]
 \centering
 \includegraphics[width=\textwidth]{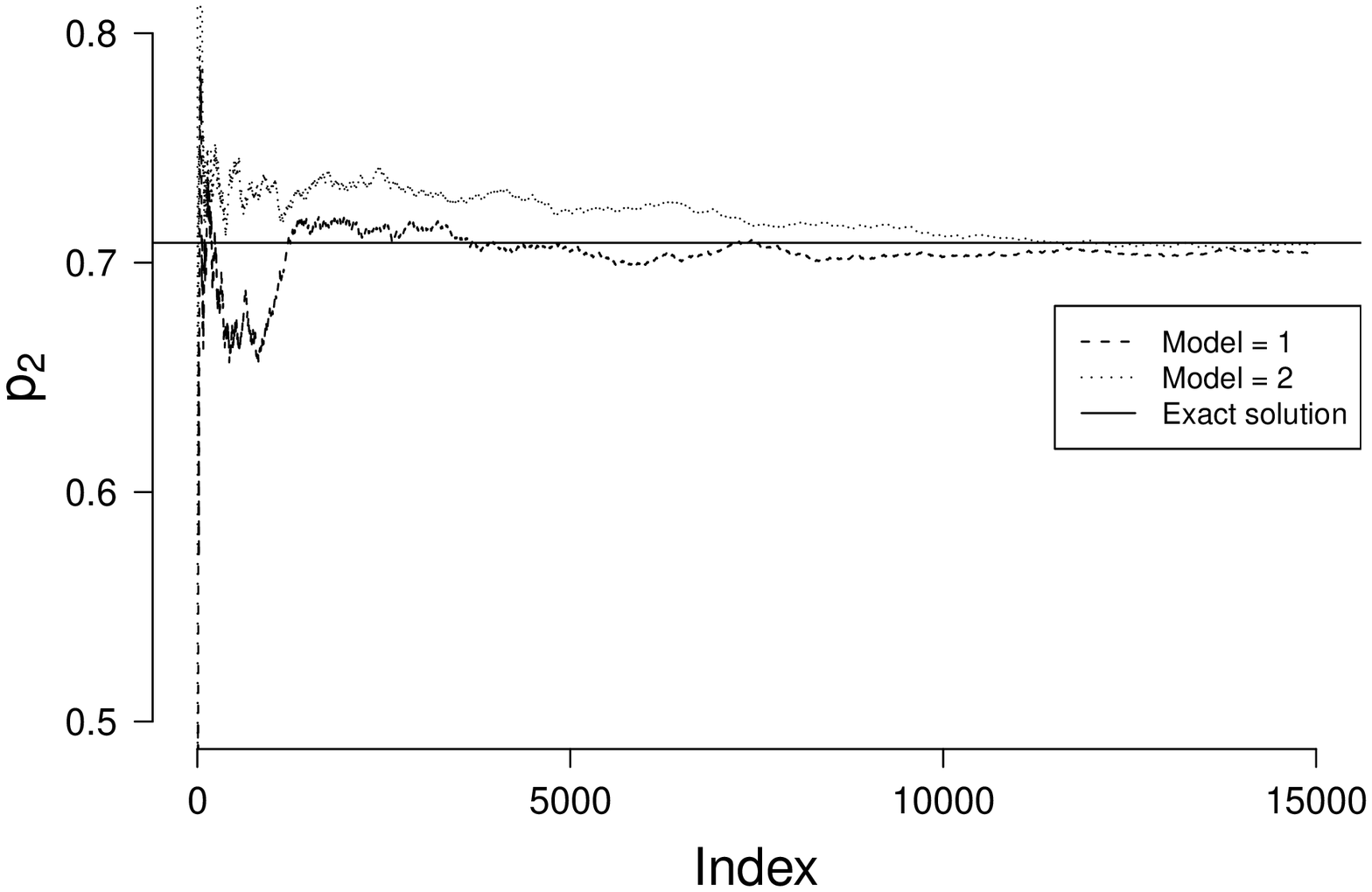}
 \caption{Index plot of the cumulative posterior probability $p = \Pr(M = 2)$ starting with model 1 (red) or model (2).  The horizontal black line corresponds to the exact result.}
 \label{fig:Pines}
\end{figure}

Following \cite{Han2001} we assigned model priors of $\Pr(M_1) = 0.9995$ and $\Pr(M_2) = 0.0005$ to ensure that the two models were visited in roughly equal proportion.  Starting at model 1 or model 2 the chain for the posterior model probability converges rapidly (Figure \ref{fig:Pines}).  After running the two chains for 200,000 iterations and discarding the first 100,000 as a burn-in, our estimate of the posterior model probability was 0.709 corresponding to a Bayes factor $BF_{21}$ of 4870.  These are in close agreement with the exact values of $\Pr(M_2 | \bm y) = 0.70865$ and $BF_{21} = 4862$ reported by \cite{Han2001}.
\begin{flushright}{\bf Figure \ref{fig:Pines} about here} \end{flushright}

Using our second method, we sampled 200,000 values of $\bm \psi$ from each chain $h$, and for the $i$th sample we calculated $\Pr(M_k | \bm \psi^{(i)}, M_h)$ $(k = 1,2)$.  Averaging across $i$ we obtain an estimated transition matrix of:
\[
 \left ( \begin{array}{cc} 0.6003 & 0.3997\\0.1651 & 0.8349 \end{array} \right )
\]
with steady-state marginal distribution of $(0.2924, 0.7076)'$, corresponding to $BF_{21} = 4838$.

\subsection{Trout return rates}
\cite{Link2006} report an analysis based on fitting logistic regression models to the return rates for brown trout expressed in terms of sex $S_i$ and length $L_i$ effects.  Modeling the return indicator $y_i \sim Bern(p_i)$ they considered five models:
 \begin{itemize}
  \item[1.] $\eta_i = \beta_0$
  \item[2.] $\eta_i = \beta_0+\beta_1S_i$
  \item[3.] $\eta_i = \beta_0+\beta_2L_i$
  \item[4.] $\eta_i = \beta_0+\beta_1S_i+\beta_2L_i$
  \item[5.] $\eta_i = \beta_0+\beta_1S_i+\beta_2L_i+\beta_3S_iL_i$
 \end{itemize}
where $\eta_i = \mbox{logit}(p_i)$.

\cite{Link2006} used the following priors on parameters:
\[
  [\bm \beta_k | V, M_k] = \left \{ \begin{array}{ll} N(0, V^{-1}) & k=1\\N(0, (2V)^{-1}) & k=2\\
                                                      N(0, (2V)^{-1}) & k=3\\
                                                      N(0, (3V)^{-1}) & k=4\\
                                                      N(0, (4V)^{-1}) & k=5 \end{array} \right .
\]
where $V$ has a $Ga(3.29, 7.80)$ prior distribution.  This choice was motivated by the observation that if logit($p$) $\sim N(0, V^{-1})$ and $V \sim \Gamma(3.29, 7.80)$, then the marginal distribution of $p$ is very nearly uniform on [0,1].  With $S_i$ and $L_i$ having been standardized, these choices of priors ensure that the prior on $e^{\eta_i}/(1-e^{\eta_i}) = p_i$ is approximately $U(0, 1)$ for $S_i = \pm 1$ and $L_i \pm 1$.  

\subsubsection*{Palette and bijections}
Each element of $\bm \psi$ is directly associated with either an element of the beta vector or with a supplemental variable $u$ (Table \ref{tab:psi2theta}):
\begin{table}[htbp]
  \centering
 \begin{tabular}{llllll}
  \hline
  &\multicolumn{5}{c}{Model}\\
  $\bm \psi$ & 1 & 2 & 3 & 4 & 5\\
  \hline
  $\psi_1$ & $\beta_0$ & $\beta_0$ & $\beta_0$ & $\beta_0$ & $\beta_0$\\
  $\psi_2$ & $u_1$     & $\beta_1$ & $u_1$     & $\beta_1$ & $\beta_1$\\
  $\psi_3$ & $u_2$     & $u_1$     & $\beta_2$ & $\beta_2$ & $\beta_2$\\
  $\psi_4$ & $u_3$     & $u_2$     & $u_2$     & $u_1$       & $\beta_{12}$\\
  \hline
 \end{tabular}
 \caption{Association between elements of $\bm \psi$ and elements of $\bm \beta_k$, specific parameters for model $M_k$ and supplemental variables $\bm u_k$ used in model $M_k$ for matching the parameter dimension to $\bm \psi$.}
 \label{tab:psi2theta}
\end{table}
The parameter $V$ is part of the prior specification and is common to all models so we chose to leave it out of the palette specification, although this is not necessary.  Updates for $V$ were stored when each model was fitted.

For a particular model, the priors on the supplemental variables were the same as the priors used for the $\beta$ coefficients in that model, and in each case the Jacobian of the transformation from $\bm \theta^{(k)}$ to $\bm \psi$ is an identity matrix of dimension 5. 

As an example, Model 1 (constant only) has parameter vector $\bm{\theta}^{(1)} = \beta_0$ with supplemental variables $\bm{u}^{(1)} =(u_1,u_2,u_3)'$.  Thus
\[
 g_1(\bm{\psi}) = g_1 \left ( \begin{array}{c} \psi_1\\ \vdots \\\psi_4 \end{array} \right ) = \bm{\Theta}^{(1)} = \left ( \begin{array}{c} \beta_0\\u_1\\u_2\\u_3 \end{array} \right ) = \left ( \begin{array}{c} \psi_1\\ \vdots \\ \psi_4 \end{array} \right ) \quad ,
\]
leading to:
\begin{eqnarray*}
 [\bm \psi | M_1, V] &=& f_1(g_1(\bm{\psi})) \left | \frac{\partial g_1(\bm{\psi})}{\partial \bm{\psi}} \right |\\
   &=& N(\psi_1;0,V^{-1}) \times N(\psi_2;0,V^{-1}) \times N(\psi_3;0,V^{-1}) \times N(\psi_4;0,V^{-1})\\
   &&\times \; Ga(V;3.29, 7.80) \times \left | \I_5 \right | .
\end{eqnarray*}
Repeating this process for each model we obtain the model-specific priors:
\[
  [\bm \psi | M_k, V] = \prod_{i=1}^4 N(\psi_i; 0, (n_kV)^{-1}) \times Ga(V; 3.29, 7.80) \times \left | \I_5 \right |
\]
where $n_k$ is the dimension of the vector $\bm \beta^{(k)}$.

For generating a chain of model indicators, we used a direct draw from the full conditional:
\[
  \Pr(M_k | \bm \psi, V) = \frac{\Pr(M_k) \prod_{i=1}^4 \sqrt{\frac{n_kV}{2 \pi}} e^{\frac{-n_kV}{2}\psi_i^2} \prod_{j=1}^{1961} p_j^{(k)}(1-p_j^{(k)})}{\sum_{h=1}^5 \Pr(M_h) \prod_{i=1}^4 \sqrt{\frac{n_hV}{2 \pi}} e^{\frac{-n_hV}{2}\psi_i^2} \prod_{j=1}^{1961} p_j^{(h)}(1-p_j^{(h)})}
\]
where $\text{logit}(p_i^{(k)}) = x_i'\bm \beta^{(k)}$.  

\begin{figure}[h!]
 \centering
 \includegraphics[width=\textwidth]{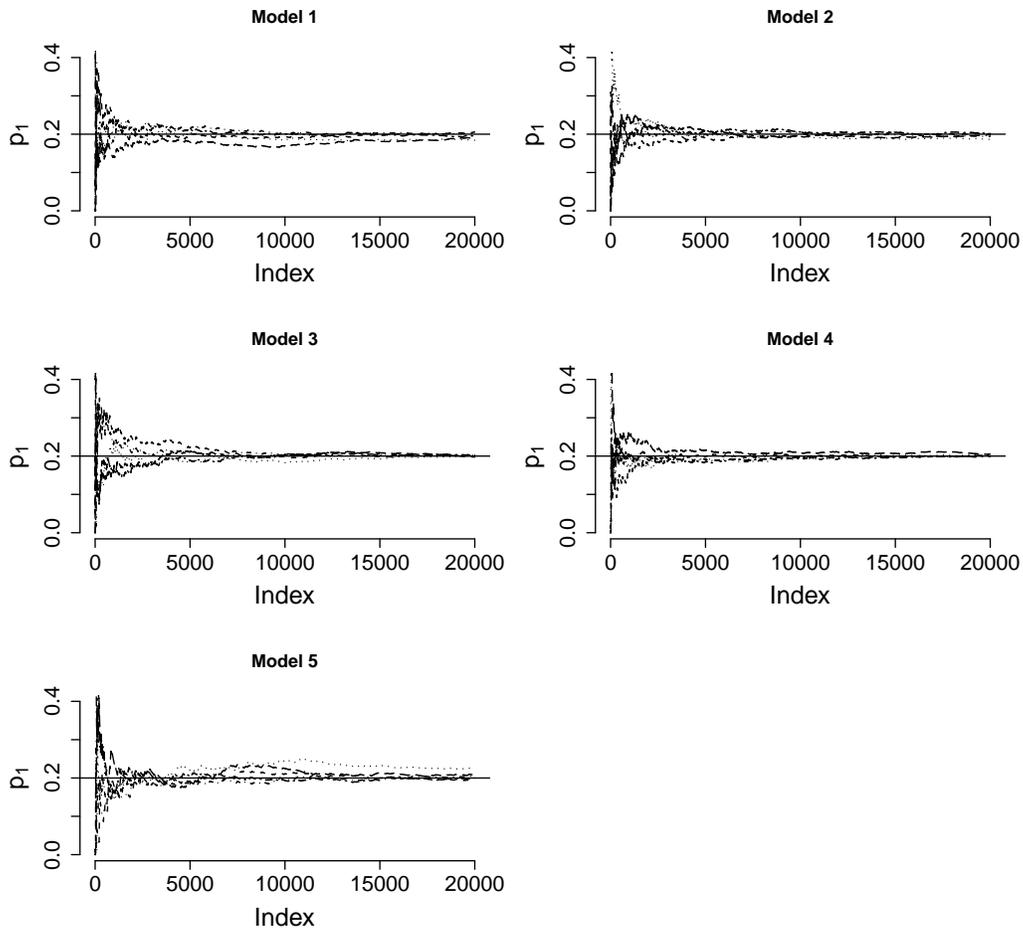}
 \caption{Index plot of the cumulative posterior model probabilities.  Each plot represents a different model probability and the different colored chains represent different starting values.  The black line corresponds to the value targeted during tuning.}
 \label{fig:Troutchains}
\end{figure}

To estimate the Bayes factors we first fitted the five models, in each case combining results from three different chains of length 500,000 after discarding a burn-in.  We then generated five chains using our Gibbs sampler for the model indicator, starting each chain with a different model.  Following \cite{Link2006} we first tuned the Gibbs sampler to visit each model in roughly equal proportion.  Mixing of the model indicators appears rapid (Figure \ref{fig:Troutchains}) and agreement with the \cite{Link2006} estimates is good after combining the results from the second half of 200,000 iterations of the five chains (Table \ref{tab:TroutBFestimates}).
\begin{table}[htbp]
  \centering
 \begin{tabular}{lll}
\hline
    $j$   & BF$_{1j}$      & $\Pr(M_j | \bm y)$\\
\hline
 1     & 1 (1)          &  0.893 (0.894)\\
 2     & 31.3 (31.7)    &  0.029 (0.028)\\
 3     & 12.3 (12.4)    &  0.073 (0.072)\\
 4     & 274.6 (281.7)  &  0.003 (0.003)\\
 5     & 383.4 ( 390.1) &  0.002 (0.002)\\
\hline
 \end{tabular}
\caption{Estimates of Bayes factors BF$_{1j}$ for comparing models 1 and $j$ and estimates of posterior model probabilities under constant prior model probabilities $\Pr(M_j) = 0.2$.  Corresponding estimates from \cite{Link2006} are given in parentheses.}
\label{tab:TroutBFestimates}
\end{table}
\begin{flushright}{\bf Figure \ref{fig:Troutchains} about here} \end{flushright}

For method two we drew a sample of 10,000 values for $\bm \psi$ from each chain \footnote{Only 10,000 samples were drawn due to the large RAM requirements on the desktop.  This number can easily be increased by writing batches of such draws to the hard-drive.} leading to an estimate of the transition matrix of:
\[
  \left ( \begin{array}{ccccc}    0.8172    &0.0870    &0.0847    &0.0088    &0.0024\\
    0.0858    &0.8086    &0.0107    &0.0755    &0.0195\\
    0.0854    &0.0102    &0.8233    &0.0759    &0.0052\\
    0.0081    &0.0749    &0.0781    &0.7884    &0.0504\\
    0.0026    &0.0176    &0.0057    &0.0498    &0.9244 \end{array} \right )
\]
with steady-state marginal distribution
\[
  \left ( \begin{array}{c} 0.1986\\
    0.1975\\
    0.2016\\
    0.1989\\
    0.2034 \end{array} \right ).
\]

\subsection{Simple binomial}
In both of the above examples, the bijections from $\bm \psi$ to $\bm \theta$ are simple 1-1 mappings with the Jacobian of the transformation an identity matrix.  Now consider an example where $Y_i \sim B(N_i, p_i)$ and we have observations $y_1 = 8$, $n_1 = 20$, $y_2 = 16$, and $n_2 = 30$.  What is the evidence for $p_1 \neq p_2$ against $p_1 = p_2 = \pi$?  To compute an appropriate Bayes factor we fit two models:
\begin{itemize}
   \item[1.] Model 1: $(p_1, p_2)$ with independent $Be(\alpha_p,\beta_p)$ priors
   \item[2.] Model 2: $p_1 = p_2 = \pi$ with a $Be(\alpha_\pi,\beta_\pi)$ prior on $\pi$.
\end{itemize}
For model 1, we assign $\bm \psi =  (p_1, p_2)'$.  It seems natural in moving from model 1 to model 2 that the average $\bar{\psi} = (\psi_1+\psi_2)/2$ should provide a good candidate for $\pi$.  
Thus, our bijections can be written as:  
\begin{eqnarray*}
   \mbox{Model 1:} & \I_2 \times \bm \psi = \left (\begin{array}{c} p_1\\ p_2 \end{array} \right )
\end{eqnarray*}
and
\begin{eqnarray*}
  \mbox{Model 2:} & \left ( \begin{array}{cc} 1/2 & 1/2 \\ 0 & 1 \end{array} \right ) \times  \left ( \begin{array}{c}\psi_1 \\ \psi_2 \end{array} \right ) = \left ( \begin{array}{c} \pi \\ u \end{array} \right ),
\end{eqnarray*}
where $\I_2$ is a $2 \times 2$ identity matrix and $u$ an appropriate supplemental variable

\begin{figure}[!htbp]
 \centering
 \includegraphics[width=\textwidth]{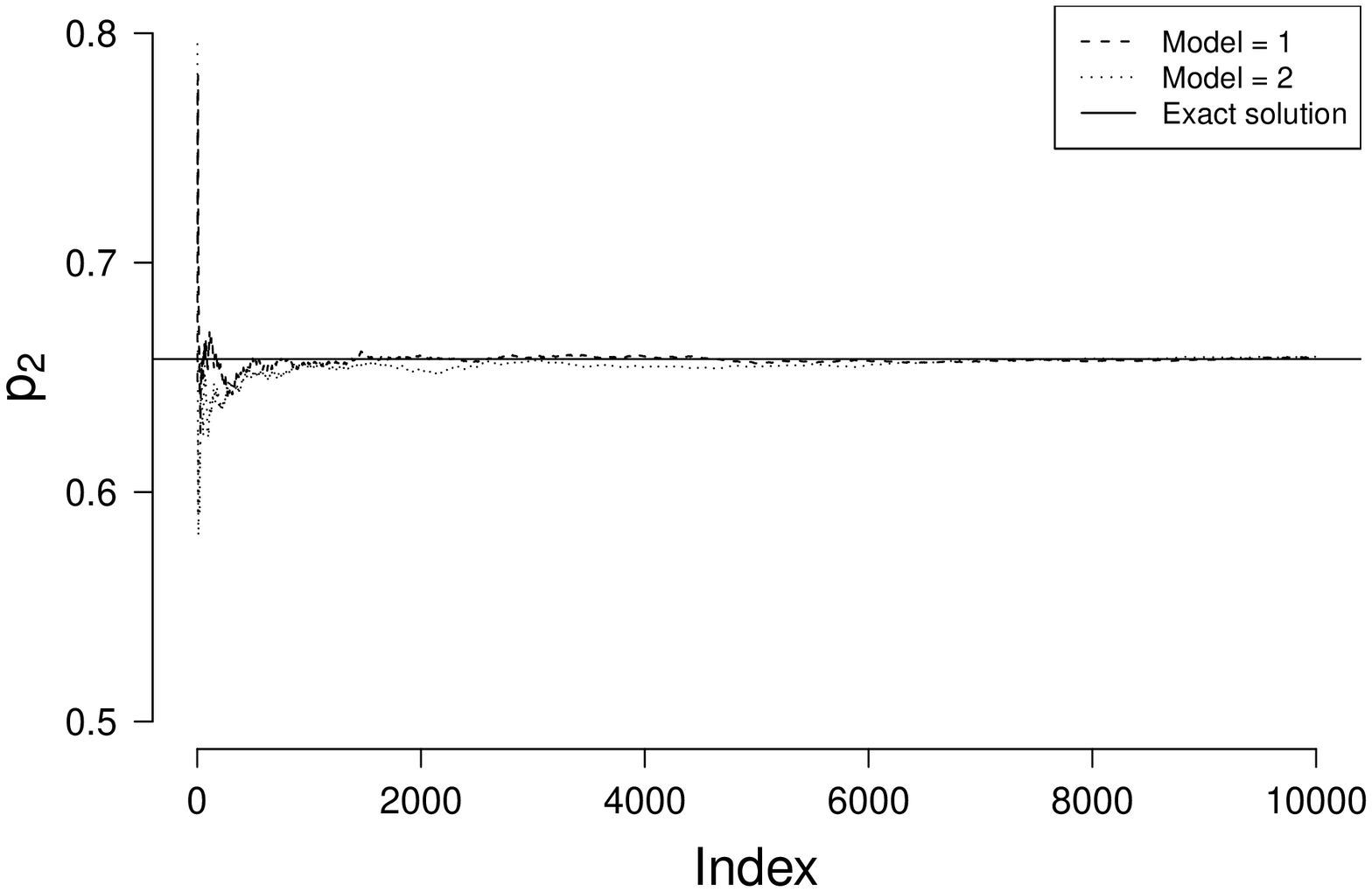}
 \caption{Index plot of the cumulative posterior probability $p = \Pr(M = 2)$ starting with model 1 (red) or model (2).  The horizontal black line corresponds to the exact result.}
 \label{fig:Binomial}
\end{figure}

Our Gibbs sampler then proceeds as follows:
\begin{itemize}
 \item[1.] Within models the full conditional distributions for model-specific parameters are of known form since we have conditional (on the model) conjugacy:
 \begin{itemize}
   \item[-] Under Model 1 we sample $p_1 \sim Be(8+\alpha_p, 12+\beta_p)$ and $p_2 \sim Be(16+\alpha_p, 14+\beta_p)$ and then compute $\bm \psi = (p_1,p_2)'$.
   \item[-] Under Model 2 we sample $\pi \sim Be(24+\alpha_\pi, 26+\beta_\pi)$ and $u \sim Be(\alpha_u,\beta_u)$ and then set $\psi_1 = 2\pi-u$ and $\psi_2 = u$.
 \end{itemize}
 \item[2.] Between models we set:
 \begin{itemize}
  \item[-] $\Pr(M_1 | \cdot) \propto \Pr(M_1) \times \psi_1^{8+\alpha_p}(1-\psi_1)^{12+\beta_p} \psi_2^{16+\alpha_p}(1-\psi_2)^{14+\beta_p} \Ind_{\psi_1 \in (0,1)} \Ind_{\psi_2 \in (0,1)}$
  \item[-] $\Pr(M_2 | \cdot)  \propto \Pr(M_2) \times \bar{\psi}^{24+\alpha_\pi}(1-\bar{\psi})^{26+\beta_\pi} \Ind_{\bar{\psi} \in (0,1)} \Ind_{\psi_2 \in (0,1)} \times \frac{1}{2}$
 \end{itemize}
 where $\Ind_E$ denotes the indicator of the event $E$ and the proportionality constant is the same for each model.  We then sample the model indicator by a direct draw from a categorical distribution with sample space $\{1,2\}$ and parameter vector $(\pi_1, \pi_2)'$ where 
 \[
    \pi_j = \frac{\Pr(M_j | \cdot)}{\Pr(M_1 | \cdot)+\Pr(M_2 | \cdot)}.
 \]
 \end{itemize}
To fit the models we used as prior parameters $\alpha_p=\beta_p=\alpha_\pi=\beta_\pi=1$ (i.e., independent $U(0,1)$ priors).  We also set $\alpha_u=\beta_u=15$ so that draws for $u$ were similar to draws for $p_2$.  Convergence of the chain for the posterior probability of model 2 was rapid (Figure \ref{fig:Binomial}). Combining results from 100,000 iterations of the two chains we obtained $\hat{B}_{21} = 1.92$ and $\Pr(M_2) = 0.658$.  For both models the marginal distribution of the data is straight-forward to compute and the exact solution for the posterior model probabilty is 0.6580.
\begin{flushright}{\bf Figure \ref{fig:Binomial} about here} \end{flushright}

Using method two with a sample of 100,000 values of $\bm \psi$ from each chain we estimate the transition matrix as:
\[
  \left ( \begin{array}{cc} 0.4318 & 0.5682\\
0.2951 & 0.7049 \end{array} \right )
\]
with steady-state marginal distribution $(0.3423 0.6577)'$.

\section{Discussion}
%
Bayesian inference offers an appealing framework for multimodel inference but the difficulties of computing Bayes factors, or equivalently posterior model probabilities, can be a barrier to implementation.  Being able to independently fit models and then post-process them using RJMCMC as we have described here offers a partial solution to the problem.  

An issue often raised in objection to Bayesian multimodel inference (BMI) based on Bayes factors is that one must assume that the true model is in the model set.  As we have argued elsewhere \cite[]{Link2006,Link2010}, this is a red-herring -- conditioning on a model set is no less innocuous than conditioning on a model as must be done for any form of statistical inference.  Conditioning on models and model sets is done for operational convenience - we no more believe that truth is in our model set than we believe that the model $y_i \stackrel{iid}{\sim} N(\mu, \sigma^2)$ can ever be a true and complete representation of any set of data.

A more serious issue with BMI is priors on parameters; it is well-known that Bayes factors are sensitive to choice of priors, particularly vague priors.  Our view is that priors should be chosen so that common features of interest in each model have the same prior uncertainty associated with them.  An attempt at such an approach is illustrated by the West Coast trout example in which priors were constructed based on the logit of the return probability for trout that had typical values of the covariates.  Such an approach we have previously referred to as ``nonpreferential'' \cite[]{Link2008a}.  

Choice of efficient bijections for moving between models requires some thought.  Although our approach simplifies this problem to one of choosing $K$ such bijections, choices must be made.  Features that are of interest and common across models can be exploited in choosing bijections as well as providing a basis for constructing non-preferential priors.  Generalized linear model formulations such 
as represented in our trout example offers one means for constructing bijections.
 One area of possible fruitful investigation in this context is that our palette representation of RJMCMC appears to be connected to the use of importance link functions \cite{Maceachern2000}.  There may be benefits from considering this connection from the point of view of determining transformations $g(\bm \psi)$ in our representation that lead to more efficient Monte Carlo estimation of posterior model probabilities. 

Our description of RJMCMC as simple Gibbs sampling with a direct draw from a known distribution for model probabilities is a further useful simplification.  Moves can be made to any model in the set ${\mathcal M}$ using samples from the full-conditional distribution for model indicators; we are not restricted to moves between pairs of models.  Methods that involve moves to neighbours have been used to automate search across very high dimensional model space.  We are skeptical about the value of such algorithms as they induce a particular prior on parameters.  Such default constructions may lead to priors that are prejudicial in which case posterior model probabilities would be more a reflection of these prior prejudices than data-informed posterior weighting.


\bibliographystyle{asa}
\bibliography{BMIrefs}
\newpage

\newpage

\newpage

\end{document}